\documentclass[preprint2]{aastex}

\usepackage{amsfonts,amssymb,bm,graphicx,natbib,color,array}

\shorttitle{Time Delay in Relativistic MOND} \shortauthors{Tian et al.}

\begin{document}

\title{\textbf{Hubble Constant, Lensing, and Time Delay in Relativistic MOND}}
\author{Yong Tian$^{1}$, Chung-Ming Ko$^{2}$, Mu-Chen Chiu$^{3}$}
\affil{$^{1}$Department of Physics, National Central University,
Jhongli, Taiwan 320, R.O.C.}\email{yonngtian@gmail.com}
\affil{$^{2}$Institute of Astronomy, Department of Physics and
Center for Complex Systems, National Central University,
Jhongli, Taiwan 320, R.O.C.}\email{cmko@astro.ncu.edu.tw}
\affil{$^{3}$Scottish University Physics Alliance, Institute for
Astronomy, the Royal Observatory,
University of Edinburgh, Blackford Hill, Edinburgh, EH9 3HJ, UK}\email{mcc@roe.ac.uk}

    \begin{abstract}

    Time delay in galaxy gravitational lensing systems has been used to determine the value of Hubble constant.
    As in other dynamical phenomena at the scale of galaxy, dark matter is often invoked in gravitational lensing
    to account for the ``missing mass'' (the apparent discrepancy between the dynamical mass and the luminous mass).
    Alternatively, modified gravity can be used to explain the discrepancy.
    In this paper we adopt the Tensor-Vector-Scalar gravity (T$e$V$e\,$S), a relativistic version of MOdified Newtonian Dynamics (MOND),
    to study gravitational lensing phenomena and derive the formulae needed to evaluate the Hubble constant.
    We test our method on quasar lensing by elliptical galaxies in the literature.
    We focus on double-image systems with time delay measurement.
    Three candidates are suitable for our study: HE 2149-2745, FBQ J0951+2635 and SBS 0909+532.
    The Hubble constant obtained is consistent with the value used in fitting the CMB result in neutrino cosmological model.

\end{abstract}


\date{}

\maketitle
    \section{Introduction}\label{sec:intro}
        The Hubble constant $H_0$ is a long-debated quantity in cosmology for more than half a century.
        Basically, it comes from the relation between the cosmological distance and the receding velocity of galaxies, $v=H_0 d$.
        Its inverse represents the age of the universe.

        The value of $H_0$ is sensitive to the way we estimate the distance.
        Its value has been estimated by many distance-determination methods, such as Cepheids, tip of the red giant branch, maser galaxies,
        surface brightness fluctuations, Tully-Fisher relation, Type Ia supernovae, gravitational time delay,
        Sunyaev-Zel'dovich (SZ) effect, and the cosmic microwave background (CMB) \citep[for details see][]{Jackson07,Freedman10}.
        In this work, we focus on gravitational lensing and time delay.
        Obtaining $H_0$ by gravitational time delay was introduced by \cite{Refsdal64}.
        Bright variable sources are needed and Refsdal suggested supernovae.
        Subsequent works on $H_0$, however, used quasar lensing.
        Now there are 18 systems of quasar lensing with time delay measurement~\citep{Paraficz10}. 

        One advantage of using time delay to derive the Hubble constant is that it is less sensitive to cosmological models.
        Thus it provides a more direct probe of the cosmological distance~\citep{Freedman10}.
        However, there are some uncertainties in determining the mass distribution by image deflections and distortions from gravitational lensing.
        This is commonly known as ``mass sheet degeneracy''~\citep{Gorenstein88}.
        Another source of uncertainty in mass is, of course, the missing mass problem.
        Missing mass is a long standing issue.
        \cite{Oort} and \cite{Zwicky} were the first to put forward the notion of missing mass in
        our Galaxy and the Coma cluster.
        The missing mass problem was neatly confirmed by the observed flat rotation curve in spiral galaxies~\citep{Rubin, vanAlba85, Bgmn89}.
        For various aspects on the history of missing mass, the reader is referred to \citet{SandersBook}.
        Nowadays, missing mass exists in nearly all types of
        galactic systems, clusters of galaxies, large scale structure and CMB.
        In fact, the problem should be interpreted in terms of excess acceleration or gravity, i.e., there are some accelerations
        which cannot be accounted for by the luminous matter only.
        To compensate the excess acceleration one can, on the one hand, introduce dark matter into the system.
        On the other hand, one can modify Newton's law of motion or the law of gravity.
        \cite{M1} proposed the MOdified Newtonian Dynamics (MOND) to explain both the flat rotation curve and the Tully-Fisher
        relation~\citep{TullyFisher}.
        MOND asserts that when the acceleration of an object, which is under the influence of gravity only, is smaller than about
        $\mathfrak{a}_0=1.21\times10^{-10}$ m s$^{-2}$, Newton's second law of motion no longer holds.
        The acceleration of the object is not proportional to the gravitational force exerted on it.
        The proposed modification is
        \begin{equation}\label{MONDEqmu}
          \tilde\mu(|\mathbf{a}|/\mathfrak{a}_0)\mathbf{a}=-{\mathbf{\nabla}}\Phi_{\rm N}=\mathbf{a}_{\rm N}\,,
        \end{equation}
        where $\mathbf{a}$ is the acceleration of the object and $\Phi_{\rm N}$ is the Newtonian gravitational potential.
        The function $\tilde\mu(x)$ is called the interpolation function.
        It is a monotonically increasing function which connects the Newtonian and the deep MOND regimes.
        With $x=|\mathbf{a}|/\mathfrak{a}_0$, $\tilde\mu(x)\approx 1$ for $x\gg 1$ (Newtonian regime), and
        $\tilde\mu(x)\approx x$ for $x\ll 1$ (deep MOND regime).
        In later calculations, we will consider spherically symmetric model.
        It is useful to introduce the inverted interpolation function $\tilde\nu$ such that
        \begin{equation}\label{MONDEqnu}
          \mathbf{a}=-{\mathbf{\nabla}}\Phi=\tilde\nu(|\mathbf{a}_{\rm N}|/\mathfrak{a}_0)\mathbf{a}_{\rm N}\,.
        \end{equation}
        For convenience we call $\Phi_{\rm N}$ the Newtonian potential and $\Phi$ the MONDian potential.

        MOND is very successful in explaining the dynamics of galactic systems~\citep[see the review by][]{SandersMcGaugh}.
        Recently, \cite{McGaugh11a, McGaugh12} showed that MOND can perfectly explain the Tully-Fisher relation in gas rich spiral galaxies without
        invoking uncertainty parameters such as the mass-to-light ratio of galaxies.
        As usual the result created some debate~\citep[][]{Foreman12, Gnedin11, McGaugh11b}.
        Nevertheless, many consider that MOND is not quite successful on the cluster of galaxies scale~\citep[see e.g.,][]{Aguirre01, Clowe06, Angus08}.
        A recent study on the gravitational redshift of clusters of galaxies~\citep{Wojtak11} has generated some debate on whether MOND is applicable
        to the cluster scale, although it seems that MOND does not have difficulty in interpreting the data~\citep{Bekenstein12}.
        In any case, the original MOND is a non-relativistic theory and cannot be applied to relativistic phenomena such as a gravitational lens and
        cosmology. Two decades after the original proposal by \cite{M1}, \cite{TeVeS} proposed the Tensor-Vector-Scalar (T$e$V$e\,$S)
        covariant relativistic gravity theory with MONDian dynamics as its non-relativistic limit.
        Adopting T$e$V$e\,$S, \cite{ckt06} derived the corresponding strong lens equation.
        More recently \cite{Milgrom09} proposed another relativistic version of MOND called BiMOND.
        It turns out that T$e$V$e\,$S and BiMOND have identical gravitational lensing equations.

        The lens equation has been applied to some galaxy lensing data, in which the mass of the galaxies has been calculated and compared
        with population synthesis~\citep[e.g.,][]{zhaoeal06, Ferr08, ckt11}.
        In a related work, \citet{Sanders08} showed that the MONDian lensing mass is consistent with the dynamical mass deduced from the
        fundamental plane of elliptical galaxies.

        In a recent study~\citet{Ferr12} claimed that strong lens data is in conflict with the MOND paradigm.
        They found that the mass deduced by gravitational lensing is larger than the value inferred by population synthesis.
        However, uncertainties still abound.
        In their paper, for more than half of the sample (five out of nine) the lens contains more than one galaxy.
        Evaluation of the MONDian acceleration of a non-spherical mass distribution is still in its infancy.
        It is not clear how they solved the problem of arbitrary mass distribution or access the uncertainty involved in their paper.
        In fact, population synthesis depends on a range of factors and physics of the lensing galaxy, where
        uncertainties are not easy to estimate.
        For instance, the estimated mass may differ a lot if a different initial mass function (IMF) is used, e.g.,
        in \citet{Ferr12} the mass obtained by the Salpeter IMF can be a factor of two larger than the Chabrier IMF.
        How to reduce the uncertainties is not clear at this stage.

        In this work, we turn our attention to the Hubble constant.
        If the derived Hubble constant is not consistent with other independent measurements, then the theory will be in trouble.
        If the time delay of the images of the lensing system can be measured, then in addition to strong lens equation we have another
        simple method to estimate the mass of the lens.
        Unlike dark matter theory in which the mass model is adjustable, no mass is non-luminous in MOND at the galaxy scale,
        and the mass density profile can be deduced solely from the brightness distribution.
        Equating the mass obtained by the lens equation and the time delay equation
        gives a relation between the acceleration constant $\mathfrak{a}_0$ and the Hubble constant $H_0$.
        Time delay provides another test for MOND or other theories.

        In the following, we describe the lens equation and the time delay equation in MOND.
        A discussion on some limiting cases is given before applying our method to the data available in the literature.
        We find three candidates suitable for our study. They are HE 2149-275, FBQ J0951+2635 and SBS 0909+532.
        Some concluding remarks will be given at the end.

    \section{Gravitational Lensing and Time Delay in Relativistic MOND}\label{sec:lensing}

        The discussion on light deflection due to the gravitational force can be traced to Newton's {\it Opticks}.
        The modern view of light deflection is a relativistic gravitational effect, in which both the time-like and space-like part in the
        metric contribute to the deflection angle.
        Using the General theory of Relativity (GR), Einstein derived a deflection angle which is just twice the Newtonian one.
        It turns out that the deflection angle derived from T$e$V$e\,$S or BiMOND is also twice the Newtonian one
        (i.e., the same as GR)---except that the Newtonian potential is replaced by the MONDian potential.
        The deflection angle by a spherical lens in the small angle approximation can be written as~\citep{ckt06, ckt11}
        \begin{equation}\label{angle1}
          \Delta\varphi=2\int a_{\perp}{dt\over c}
          \approx\frac{ 2\varrho_0}{c^{2}}\int^{D_{\rm L}^\prime}_{-D_{\rm LS}}
          {1\over\varrho}{\partial\Phi(\varrho)\over\partial\varrho}\,d\zeta\,,
        \end{equation}
        where c is the speed of light, $\theta$ is the image position, $\varrho$ is the distance from the center of the spherical lens,
        $\varrho_0\approx D_{\rm L}\theta$ is the closest approach of the light path from the center of the lens,
        $\zeta^2=\varrho^2-\varrho_0^2$, and $\Phi(\varrho)$ is the MONDian potential.
        $D_{\rm L}$, $D_{\rm L}^\prime$ and $D_{\rm LS}$ are the angular distances of the lens from the observer,
        the observer from the lens and the source from the lens, respectively.
        The direction of the image is in the direction of the closest approach (projected on the sky).
        For a spherical lens there are two images located at both side of the source, and the governing equation (called the lens equation)
        is given by
        \begin{equation}\label{lensEq}
          \beta=\theta_+-\alpha(\theta_+)=\alpha(\theta_-)-\theta_-\,,\quad
          \alpha(\theta)=\Delta\varphi {D_{\rm LS}\over D_{\rm S}}\,,
        \end{equation}
        where $\beta$ is the source position and $\theta_\pm$ are the image positions.
        The upper sign denotes an image on the same side as the source and the lower sign on the opposite side of the source.
        $\alpha(\theta)$ is commonly called the reduced deflection angle.

        Time delay is defined as the difference in time traveled by light along the actual path and along the undeflected path.
        It can be derived from Fermat's principle or from the geodesic equation in relativistic gravitation theory.
        As with the deflection angle, the form of the time delay is the same for GR and MOND
        (with the Newtonian potential for GR and the MONDian potential for MOND),
        \begin{equation}\label{timedelay}
          t(\theta)={(1+z_{\rm L})\over c}\left[{D_{\rm L}D_{\rm S}\over 2D_{\rm LS}}\alpha(\theta)^2
          -\int^{D_{\rm L}^\prime}_{-D_{\rm LS}}\frac{2\Phi(\varrho)}{c^{2}}\,d\zeta\right]\,.
        \end{equation}
        The first and second term in Equation~(\ref{timedelay}) are referred to as the geometric and the potential time delay.
        In the cases where the difference in time delay of the two images is available, the value of $H_0$ (and the mass of the lens as well)
        can be obtained by solving the time delay difference equation (time delay equation for short)
        \begin{equation}\label{timedelayEq}
          \Delta t=t(\theta_-)-t(\theta_+)\,,
        \end{equation}
        and the lens equation Equation~(\ref{lensEq}).

        To illustrate the ideas, we start with a simple example. We consider a point mass lens in Bekenstein form.
        Bekenstein form is a frequently used interpolation function with a very simple inverted form
        $\tilde{\nu}(x_{\rm N})=1+1/\sqrt{x_{\rm N}}$ (where $x_{\rm N}=|\mathbf{a}_{\rm N}|/\mathfrak{a}_0$).
        Equation~(\ref{MONDEqnu}) becomes
        \begin{equation}\label{pointBekenstein}
          {\partial\Phi\over\partial\varrho}={\partial\Phi_{\rm N}\over\partial\varrho}
          +\sqrt{\mathfrak{a}_0{\partial\Phi_{\rm N}\over\partial\varrho}\,}\,,\quad
          {\partial\Phi_{\rm N}\over\partial\varrho}={GM\over\varrho^2}\,.
        \end{equation}
        The corresponding lens equation and time delay equation are
        \begin{equation}\label{MONDlens}
          \frac{\theta_+\theta_-}{\theta_{\rm E}^2}=1+\frac{\theta_+\theta_-}{(\theta_++\theta_-)}\frac{\pi}{\theta_0}\,,
        \end{equation}
        \begin{equation}\label{GRtimedelay}
          {{\tilde D}_{\rm LS}\over{\tilde D}_{\rm L}{\tilde D}_{\rm S}}{H_0\Delta t\over(1+z_{\rm L})}=
          {1\over 2}\left(\theta_+^2-\theta_-^2\right)+\theta_{\rm E}^2\log\left({\theta_+\over\theta_-}\right)\,,
        \end{equation}
        where
        \begin{equation}\label{thetaEtheta0}
          \theta_{\rm E}^2={4GMD_{\rm LS}\over c^2D_{\rm L}D_{\rm S}}\,,\quad
          \theta_0^2={GM\over\mathfrak{a}_0 D_{\rm L}^2}\,.
        \end{equation}
        $\theta_{\rm E}$ is called the Einstein radius.
        In Equation~(\ref{GRtimedelay}), the angular distance is normalized to $c/H_0$, i.e.,
        $D_{\rm L}=(c/H_0){\tilde D}_{\rm L}$, $D_{\rm S}=(c/H_0){\tilde D}_{\rm S}$, $D_{\rm LS}=(c/H_0){\tilde D}_{\rm LS}$.
        Moreover, ${\tilde D}_{\rm L}$, ${\tilde D}_{\rm S}$, ${\tilde D}_{\rm LS}$ depend on the redshift $z$ only and not on $H_0$.

\begin{figure*}[ht]\label{timeHubble}
        \centering
        \includegraphics[width=14cm]{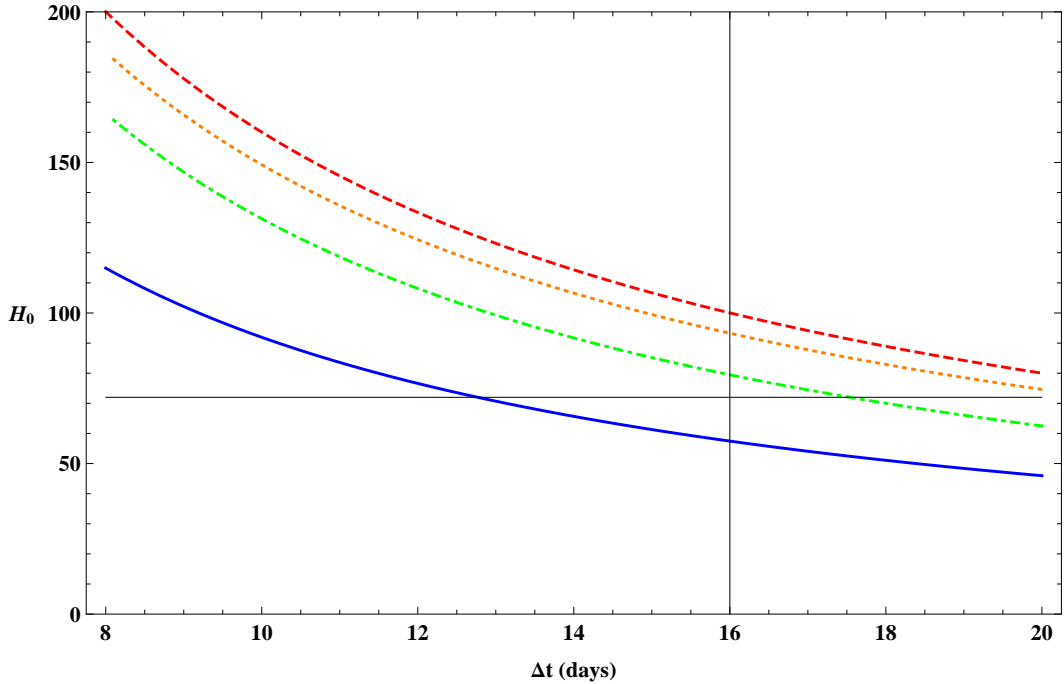}
        \caption{
        \footnotesize{
        The Hubble constant (in km s$^{-1}$ Mpc$^{-1}$) is plotted against the time delay difference (in days) of a point mass lens model and a Hernquist Model.
        We use the $\Lambda$CDM cosmology for GR, and 11 eV massive sterile neutrino cosmology for MOND.
        The dashed line is the point mass model in GR, the dotted-line the Hernquist model in GR, the dot-dashed line the Hernquist model in MOND,
        and the solid line is the point mass model in deep-MOND.
        To put the diagram into perspective, we consider the quasar lensing object FBQ J0951+2635.
        The time delay difference of this object is 16 days and is marked by the vertical black line.
        For reference, $H_0=72$ km s$^{-1}$ Mpc$^{-1}$ is marked by the horizontal black line.
        }}        
        \end{figure*}

        It is interesting to note that the time delay equation for a point mass lens in GR is exactly the same as Equation~(\ref{GRtimedelay}).
        (In fact, this is also true for a lens with Hernquist profile, which we use in next section.)
        We can use $\theta_0$ to characterize the different regimes. If the gravitational acceleration at the closest approach is much larger
        than $\mathfrak{a}_0$ (i.e., $\theta\ll \theta_0$), it is in the Newtonian regime, and if the acceleration is much smaller than
        $\mathfrak{a}_0$ (i.e., $\theta\gg \theta_0$), it is in the deep MOND regime.
        From Equation~(\ref{MONDlens}), we know that $\theta_{\rm E}^2\rightarrow\theta_+\theta_-$ as $\theta_\pm\ll \theta_0$ (Newtonian) and
        $\theta_{\rm E}^2\approx\theta_0(\theta_++\theta_-)/\pi$ as $\theta_\pm\gg \theta_0$ (MONDian).
        The deduced $\theta_{\rm E}$ (hence the mass) from MOND is smaller than that of GR (which is what MOND is designed for).
        Hence the deduced Hubble constant $H_0$ from MOND is smaller than that of GR (see Equation~(\ref{GRtimedelay})).
        In Figure~\ref{timeHubble}, we plot the relation of $H_0$ against $\Delta t$ for a point mass model in MOND and GR.
        We also plot the relation for the Hernquist model, which will be useful in next section.

        Although the time delay equations for point mass lens in GR and MOND in Bekenstein form
        have exactly the same form as shown in Equation~(\ref{GRtimedelay}), there is one subtle difference.
        In both GR and MOND, the geometric time delay $\Delta t_{\rm G}$ can be written as
        \begin{equation}\label{Geotimedelay}
          {{\tilde D}_{\rm LS}\over{\tilde D}_{\rm L}{\tilde D}_{\rm S}}{H_0\Delta t_{\rm G}\over(1+z_{\rm L})}=
          {\theta_{\rm E}^2\over 2\theta_+\theta_-}\left(\theta_+^2-\theta_-^2\right)\,.
        \end{equation}
        In GR $\theta_{\rm E}^2=\theta_+\theta_-$, hence the first term of Equation~(\ref{GRtimedelay}) is the geometric time delay in GR.
        However, because of Equation~(\ref{MONDlens}), the geometric time delay in MOND is less that the first term in Equation~(\ref{GRtimedelay}).
        The interesting fact is the potential time delay $\Delta t_{\rm P}$ in MOND (Bekenstein form for point mass and Hernquist model) is        
        \begin{equation}\label{Pottimedelay}
        \footnotesize
          \begin{array}{rl}
          {\displaystyle
          {{\tilde D}_{\rm LS}\over{\tilde D}_{\rm L}{\tilde D}_{\rm S}}{H_0\Delta t_{\rm P}\over(1+z_{\rm L})}
          }
          &
          {\displaystyle
          =\theta_{\rm E}^2\left[\log\left({\theta_+\over\theta_-}\right)+{\pi\over 2\theta_0}(\theta_+-\theta_-)\right]
          }\\
          &
          {\displaystyle
          =\theta_{\rm E}^2\left[\log\left({\theta_+\over\theta_-}\right)
          +{1\over 2}(\theta_+^2-\theta_-^2)\left({1\over\theta_{\rm E}^2}-{1\over\theta_+\theta_-}\right)\right]\,.
          }
          \end{array}
        \end{equation}
        Thus part of the potential time delay cancels the geometric time delay ``exactly'', and renders the total time delay to have
        the same form as in GR.
        Therefore, the terms in Equation~(\ref{GRtimedelay}) have a clear identification, namely, the first term is the geometric time delay
        and the second is the potential time delay, but it is not so in MOND, where the first term of Equation~(\ref{GRtimedelay}) is partly
        geometric and partly potential, and the second term is part of the potential time delay.
        Moreover, when one approaches the deep MOND regime ($\theta_0/\theta_\pm\rightarrow 0$), $\theta_{\rm E}\rightarrow 0$ and the
        MONDian geometric time delay and the second term in Equation~(\ref{GRtimedelay}) tend to zero.
        The time delay becomes solely potential time delay.

        In the deep MOND regime, the interpolation function becomes $\tilde{\mu}(a/\mathfrak{a}_{0})\simeq a/\mathfrak{a}_{0}$.
        If the extent of the luminous matter is also much smaller than $\theta$ (i.e., can be modeled practically by a point mass),
        then the time delay difference is solely determined by the potential time delay,
        because the deflection angle approaches a constant in the deep MOND regime~\citep{ckt06}.
        In this case, the time delay (difference) equation is independent of the choice of interpolation function,
        \begin{equation}\label{deepMONDtimedelay}
          {{\tilde D}_{\rm LS}\over{\tilde D}_{\rm L}{\tilde D}_{\rm S}}{H_0\Delta t\over(1+z_{\rm L})}=
          {1\over 2}\left(\theta_+^2-\theta_-^2\right)\,.
        \end{equation}
        We emphasize that Equation~(\ref{deepMONDtimedelay}) does not have any free parameters for the interpolation
        function or the mass of the lens, not even $\mathfrak{a}_{0}$ (the utmost important constant of MOND).
        $\mathfrak{a}_{0}$ and the mass of lens are hidden in the lens equation, Equation~(\ref{MONDlens}).

        We point out that Equation~(\ref{deepMONDtimedelay}) is identical to GR with an isothermal lens model~\citep[see][]{Witt00}.
        This is expected as both potentials have the same form, namely, a logarithmic potential.
        However, Equation~(\ref{deepMONDtimedelay}) in MOND is valid only in the deep MOND regime.
        Now if some data satisfies Equation~(\ref{deepMONDtimedelay}) and is in the Newtonian regime ($\theta\ll \theta_0$),
        this will constitute a possible falsification of relativistic MOND, at least in the simple formulation we presented here.

    \section{Data and Modeling}\label{sec:data}

        Although hundreds of examples of quasar lensing have been found, only a few have been measured with gravitational time delay.
        This method has been proven difficult because the amplitude of quasar variability is quite small,
        A clear and simple modeled relative image is hard to identify~\citep[see e.g.,][]{Freedman10}.
        As far as we know only 18 strong lenses have time delay measured~\citep{Paraficz10}.
        To test our theory, we select elliptical galaxy lensing systems with double images.
        Only 5 cases satisfy our criteria. They are HE 2149-275, FBQ J0951+2635, SBS 0909+532, SDSS J1650+4251 and HE 1104-1805.
        The rest are clusters, spiral galaxies, multiple images or multiple lens, which require an analysis beyond the scope of this paper
        and involve more uncertainty.
        Moreover, the lens galaxy in SDSS J1650+4251 is very dark. It does not have a reliable effective radius.
        The uncertainty in time delay measurement in HE 1104-1805 is too large.
        This brings us to a total of three candidates: HE 2149-275~\citep{Wisotzki93,Burud02}, FBQ J0951+2635~\citep{White00, Jakobsson05},
        SBS 0909+532~\citep{Lubin00, Ullan05}.
        Table~\ref{sample} lists some properties of these three selected systems.


        \begin{table}[ht]
        \caption{
        \footnotesize{
        Data of the three selected quasar lenses with time delay difference measurement. The unit of $\theta_{\pm}$ is arcsecond.
        Effective radius in H band is used for the objects HE 2149-2745, FBQ J0951+2635~\citep{Kochanek00}, and SBS 0909+532~\citep{Lehar00}.
        Ellipticity $\epsilon=1-b/a$ is measured in the R band~\citep{Lopez98, Jakobsson05}. 
        }
        }
        \tabcolsep=2pt
        \footnotesize
        \begin{tabular}{c|ccccccc}
        \hline\hline
        Name& $z_{l}$ & $z_{s}$ & $\theta_{+}''$ & $\theta_{-}''$ & $\theta_{\rm eff}''$ & $\epsilon$ & $\Delta t$(days) \\
        \hline
        HE2149-2745    &   0.495   &   2.030    &   1.354    &   0.344   &   0.501  & 0.5   & 103$\pm$12\\
        FBQJ0951+2635  &   0.240   &   1.246    &   0.879    &   0.221   &   0.166  & 0.25  & 16$\pm$2\\
        SBS0909+532    &   0.830   &   1.376    &   0.756    &   0.415   &   1.580  &  -    & 45$^{+1}_{-11}$\\        
        \hline
        \end{tabular}
        \label{sample}
        \end{table}

        A rough estimation gives that the gravitational acceleration of the three systems ranges from $10^{-9}$ m s$^{-2}$ to $10^{-10}$ m s$^{-2}$.
        Thus these gravitational lenses are not in the deep MOND regime.
        We need to pick an interpolation function. In the literature several practical forms have been used,
        such as, the Bekenstein form $\tilde{\mu}(x)=(-1+\sqrt{1+4x})/(1+\sqrt{1+4x})$~\citep{TeVeS},
        the simple form $\tilde{\mu}(x)=x/(1+x)$~\citep{Famaey05}, and the standard form $\tilde{\mu}(x)=x/{\sqrt{1+x^2}}$~\citep{M1}.
        Recently, a theoretical form derived from quantum effects in an accelerating universe was proposed
        $\tilde{\mu}(x)=(\sqrt{4x^2+1}-1)/2x$~\citep{Ho10}.
        All these forms can be put into a canonical form~\citep{ckt11}.
        The corresponding inverted canonical interpolation function is
        \begin{equation}\label{inverted}
          \tilde{\nu}(x_{\rm N})=\left[1+{1\over 2}\left(\sqrt{4x_{\rm N}^{-\alpha}+\eta^2}-\eta\right)\right]^{1/\alpha}\,.
        \end{equation}
        In the following calculations we take the Bekenstein form, i.e., $(\alpha,\eta)=(1,0)$.

        The surface brightness profile of the lens elliptical galaxies satisfies the de Vaucouleurs' profile.
        We, therefore, adopt the Hernquist mass density profile~\citep{hern90}.
        The Newtonian gravitational potential and acceleration of the Hernquist model are
        \begin{equation}\label{Hernquist}
          \Phi_{\rm N}=-{GM\over(\varrho+\varrho_{\rm h})}\,,\quad
          {\partial\Phi_{\rm N}\over\partial\varrho}={GM\over(\varrho+\varrho_{\rm h})^2}\,,
        \end{equation}
        where the Hernquist radius $\varrho_{\rm h}$ is 0.551 times the effective radius (or half-light radius).
        The gradient of the MONDian potential is given by Equation~(\ref{MONDEqnu}).
        The lens equation and the time delay equation for the Hernquist lens in Bekenstein form are described in Appendix~\ref{sec:HernquistEq}.

        Different cosmological models give a different angular distance for the same redshift.
        MOND has been criticized that it can not form large-scale structure.
        Basically, the criticism originated from an argument in GR with baryons only.
        However, the non-linear growth of structure in MOND with neutrinos can reproduce the power spectrum ~\citep{Skordis06,Skordis09,Angus09,Diaferio12}.
        Although T$e$V$e\,$S fields has small contribution ($\sim10^-3$ or less) to the background FLRW equation~\citep{Skordis09}, this can make it affect non-linear growth in large-scale structure but close to GR in cosmology such as CMB.

        \citet{Skordis06} showed that in order to comply with the CMB observations, T$e$V$e\,$S needed 
        $(\Omega_{\rm B},\Omega_{\nu},\Omega_{\Lambda})=(0.05,0.17,0.78)$ and 2 eV massive neutrinos
        (treated as non-relativistic particles).
        However, in this model, the predicted third acoustic peak of CMB is lower than what is observed.
        \citet{McGaugh99} proposed a simple Ansatz for CMB problem in MOND: any relativistic MOND theory should contain GR in the appropriate strong-field limit. \citet{Angus09} estimated the gravity before recombination is pretty strong than MOND effect around $570\mathfrak{a}_0$.

        Then, \citet{Angus09} proposed a model with 11 eV sterile neutrinos and got a better fit to the third peak.
        (11 eV sterile neutrino is consistent with the analysis of Miniboone experiment
        which gave the mass range of sterile neutrinos as 4 eV to 18 eV ~\citet{Giunti08}).
        The sterile neutrino cosmological model of \citet{Angus09}
        is $(\Omega_{\rm B},\Omega_{\nu_{s}},\Omega_{\Lambda})=(0.05,0.23,0.72)$ and $m_{\nu_{s}}=11eV$.
        We use both 2 eV neutrino model and 11 eV sterile neutrino model as our cosmological model for angular distance calculation.
        In the three cases we studied, the difference between the two models is small.
        The difference in $\tilde{D}_{LS}/(\tilde{D}_{L}\tilde{D}_{S})$ is less than 0.4\%.

    \section{Result and Conclusion}\label{sec:result}

        Basically, our model comprises three parts: (1) a galaxy, (2) lensing, and (3) a gravity theory.
        (1) We model the lensing galaxy by Hernquist's model. Photometric measurement of the galaxy could give the effective radius
        (or half light radius) of the galaxy, and the Hernquist radius is related to the effective radius.
        The only unknown is the total mass of the lensing galaxy, $M$.
        (2) On the part of gravitational lensing, we consider strong lensing.
        The angular position and the redshift (of the lens and the two images) and the time delay difference between the two images are measured.
        When a cosmological model is adopted the only unknown is the Hubble constant $H_0$.
        (3) We consider MOND in the Bekenstein form. The only unknown is the acceleration constant $\mathfrak{a}_0$.
        These three unknowns are constrained by the lens equation and time delay equation (Equations~(\ref{HernLensEq}) \& (\ref{Hernquist})).
        In this paper, we would like to assume a gravity theory and find the mass and Hubble constant.
        The flat rotation curve of spiral galaxies and the Tully-Fisher relation in gas rich galaxies give a consistent value of the
        acceleration constant $\mathfrak{a}_0=1.21\times10^{-10}$ m s$^{-2}$~\citep{SandersMcGaugh, Famaey05, McGaugh11a}.
        We evaluate $H_0$ by Equation~(\ref{a0H0}) and $M$ by Equation~(\ref{Hernmass}).
        The results for the three selected systems (HE 2149-2745, FBQ J0951+263, SBS 0909+532) are summarized in Table~\ref{Result}.

        \begin{table*}[ht]
        \caption{
        \footnotesize{The evaluated mass of the lens and Hubble constant. The unit of mass is $10^{10}$ $M_\odot$
        and the unit of the Hubble constant is km s$^{-1}$ Mpc$^{-1}$. $x=a/\mathfrak{a}_0$ is a measure of the MONDian regime
        at the closest approach $\varrho_{0}$, and $x_\pm$ correspond to $\theta_\pm$. The smaller numbers in the first four columns are
        results taking into account the corresponding upper and lower uncertainties. We take $\Lambda CDM$ to be $(\Omega_{\rm B},\Omega_{\nu_m},\Omega_{\Lambda})=(0.05,0.23,0.72)$,  $11eV \nu_{s}$ cosmological model to be $(\Omega_{\rm B},\Omega_{\nu_{s}},\Omega_{\Lambda})=(0.05,0.23,0.72)$, and $2eV \nu$ cosmological model to be $(\Omega_{\rm B},\Omega_{\nu},\Omega_{\Lambda})=(0.05,0.17,0.78)$
        The result of GR with isothermal model is identical to MOND with point mass model.
        }}
        \tabcolsep=4pt
        \extrarowheight=3pt
        \small
        \begin{tabular}{@{}c|cc|cccc|cc@{}}
        \hline\hline
         & Mass & & $H_0$ & & &   \\
        \hline
         & GR & MOND & GR & GR & MOND & MOND & $x_-$ & $x_+$ \\
        Mass Model & Hernquist & Hernquist & Hernquist & Isothermal & Hernquist & Hernquist & Hernquist & Hernquist \\
        \hline
        Cosmology & $\Lambda CDM$ & $11eV \nu_{s}$ & $\Lambda CDM$ & $\Lambda CDM$ & $11eV \nu_{s}$ & $2eV \nu$ & $11eV \nu_{s}$ & $11eV \nu_{s}$ \\
        \hline
        HE 2149-2745	&	$	23.2 	^{	25.9 	}_{	20.5 	}$ & $	16.2 	^{	17.7 	}_{	14.7 	}$ & $	72.2 	_{	64.7 	}^{	81.8 	}$ & $	 47.7 	
        ^{	54.0 	}_{	42.7 	}$ & $	57.7 	^{	 66.1 	}_{	51.1 	}$ & $	57.6 	^{	66.0 	}_{	51.0 	}$ & 	11.9 	&	2.4 	\\
        FBQS J0951+2635	&	$	2.9 	^{	3.3 	}_{	2.6 	}$ & $	2.3 	^{	2.5 	}_{	2.1 	}$ & $	93.3 	_{	82.9 	}^{	106.6 	}$ & $	 57.5 	
        ^{	65.7 	}_{	51.1 	}$ & $	79.5 	^{	 91.7 	}_{	70.0 	}$ & $	79.3 	^{	91.5 	}_{	69.9 	}$ & 	29.2 	&	4.1 	\\
        SBS 0909+523	&	$	77.2 	^{	78.9 	}_{	58.3 	}$ & $	56.5 	^{	57.5 	}_{	44.7 	}$ & $	84.6 	_{	82.7 	}^{	112.0 	}$ & $	 81.9 	
        ^{	108.4 	}_{	80.2 	}$ & $	70.4 	^{	 95.3 	}_{	68.8 	}$ & $	70.2 	^{	95.1 	}_{	68.6 	}$ & 	9.4 	&	6.3 	\\
        \hline
        \end{tabular}
        \label{Result}
        \end{table*}

        In the two last columns of Table~\ref{Result} $x=a/\mathfrak{a}_0$
        is the ratio of the acceleration at the closest approach to the MOND acceleration constant.
        Recall that $x\gg 1$ is the Newtonian regime and $x\ll 1$ is the deep MOND regime.
        In Table~\ref{Result}, we see that the deep MOND point mass model did not give a reasonable value of $H_{0}$.
        This is understandable because these three cases are not in the deep MOND regime.
        In addition, the closest approach distance of the three cases is comparable to
        the measured Hernquist radius, thus the point mass model is not a good approximation.
        Recall that the isothermal mass model in GR and the deep MOND point mass model have an identical expression, Equation~(\ref{deepMONDtimedelay}).
        The two would give the same result (give and take for some slight difference due to the different cosmological models).
        Perhaps it is not a coincidence that these three cases are not as successful as other time delay cases by the simple isothermal model
        in GR as shown in~\citet{Witt00}.

        In general, MOND gives a smaller mass than GR, and the excess mass in GR can be interpreted as missing mass or dark matter.
        In the Newtonian regime ($x\gg 1$), MOND and GR should give a similar mass.
        For the three cases in Table~\ref{Result}, mostly $x$ is of the order of 1 to 10 (intermediate MOND regime).
        As expected the mass and Hubble constant from MOND are smaller than the value computed in GR without dark matter.
        Once again, as in other galactic scale dynamical phenomena, MOND gives a consistent picture of explaining the observed excess acceleration
        in gravitational lensing including the time delay phenomenon.

        A source of uncertainty is the choice of the interpolation function.
        Recall that either in the Newtonian regime ($x\gg 1$) or in the deep MOND regime ($x\ll 1$), different interpolation functions should
        give the same result.
        However, our sample lies in the intermediate MOND regime.
        Table~\ref{Result} shows the result from the Bekenstein form.
        Other interpolation functions are expected to give somewhat different result.
        In any case, the major uncertainty comes from observation, in particular, the time delay measurement.

        The Hubble constant obtained from lensing and time delay, of course, must be consistent with the values from other measurements.
        The Hubble constant sets the scale of distance.
        Hubble's law relates the distance of an object to its redshift.
        Recently \citet{Riess09,Riess11} calibrated low redshift type Ia supernovae with Cepheids and obtained a Hubble constant of
        $H_0=73.8 \pm 2.4$ km s$^{-1}$ Mpc$^{-1}$.
        Neutrino cosmological models will not change this value since they also give the same linear Hubble's law at low redshifts.
        The $H_0$ found by time delay in this work (see Table~\ref{Result}) is consistent with the value(s) by type Ia supernova data.

        In \citet{Angus09}, neutrino cosmological models in T$e$V$e\,$S were considered to fit the CMD acoustic spectrum.
        A 11 eV sterile neutrino model can fit the data well.
        The parameters used in the model were $H_0=71.5$ km s$^{-1}$ Mpc$^{-1}$,
        $(\Omega_{\rm B},\Omega_{\nu_{s}},\Omega_{\Lambda})=(0.05,0.23,0.72)$.
        Our result on $H_0$ is consistency with this (see Table~\ref{Result}).

        In summary, this work is a first attempt to use MOND to interpret data from gravitational time delay.
        The Hubble constant evaluated for the sample in this study is consistent with the value
        obtained from Hubble's law and also those in the literature (see Table~\ref{Result}).
        When compared with GR in the Hernquist model (without dark matter component), the evaluated mass of the lens in MOND
        is 27\% to 43\% smaller than those from GR, and the Hubble constant is 17\% to 25\% smaller than GR.

        Applying a gravity theory to a static distribution of mass gives the dynamics (or equation of motion)
        of a point mass or a photon in the mass distribution.
        It is expected that some of the parameters in the dynamics are related to the mass distribution and some of them
        are unique to the underlying gravity theory.
        In principle, the parameters can be fixed by measuring the dynamics.
        Similarly, when we apply the gravity theory to the dynamics of the universe, some parameters are related to the
        energy content of the universe and some to the theory itself.

        In the case of strong gravitational lensing, if we can assume or measure the density profile of the mass distribution
        (in MOND, the profile is supposed to be given by the brightness distribution),
        then there remains only one unknown parameter for the mass distribution, namely, the mass scale (e.g., the total mass).
        If both the image positions and the time delay are measured, then we can get rid of the mass scale.
        The remaining parameters in the dynamics are related directly (or through a distance scale)
        to the gravity theory. For instance, in MOND the remaining parameters are the acceleration constant $\mathfrak{a}_0$
        and the Hubble constant $H_0$. In this article we adopted $\mathfrak{a}_0$ from other measurements
        \citep{SandersMcGaugh, Famaey05, McGaugh11a}, and the derived $H_0$ is consistent with the value from Hubble's law
        measured by type Ia supernovae and the value needed for fitting the CMB acoustic spectrum by neutrino cosmological model.
        Gravitational lensing promises to provide a testing ground for modified gravity.

\acknowledgements
We are grateful to J.M. Nester and C. Skordis for helpful discussion and T.H. Peng for data collection.
We thank the anonymous referee for many valuable comments on an earlier version of this paper.
This work is supported in part by the Taiwan National Science Council
Grants NSC 98-2923-M-008-01-MY3 and NSC 99-2112-M-008-015-MY3.

\appendix

\section{Lens and Time Delay Equations for the Hernquist Model in Bekenstein Form}\label{sec:HernquistEq}

      In this appendix we write down the equations necessary for the analysis in Sections~\ref{sec:data} \& \ref{sec:result}.
      The lens equation for a spherical Hernquist lens (see Equation~(\ref{Hernquist})) in Bekenstein form
      ($\alpha=1$, $\eta=0$ in Equation~(\ref{inverted})) can be written as (cf. Equation~(\ref{MONDlens}))
      \begin{equation}\label{HernLensEq}
        {(\theta_++\theta_-)\over(f_++f_-)\theta_{\rm E}^2}
        =1+{(g_++g_-)\over(f_++f_-)\theta_0}\,,
      \end{equation}
      where $f_\pm=f(\theta_\pm,\theta_{\rm h})$, $g_\pm=g(\theta_\pm,\theta_{\rm h})$
      \begin{equation}\label{fg}
        f(\theta,\theta_{\rm h})={\theta\left[1-\theta_{\rm h}h(\theta,\theta_{\rm h})\right]
        \over(\theta^2-\theta_{\rm h}^2)}\,,\quad
        g(\theta,\theta_{\rm h})=\theta h(\theta,\theta_{\rm h})\,,
      \end{equation}
      \begin{eqnarray}\label{h}
        h(\theta,\theta_{\rm h})
          & =\left\{\matrix{{\displaystyle {1\over\sqrt{\theta^2-\theta_{\rm h}^2\,}}
                    \left[{\pi\over 2}-\sin^{-1}\left(\theta_{\rm h}\over\theta\right)\right]\,,}
                    & {\rm for}\ \theta_{\rm h}<\theta \cr
                    {\displaystyle 1\,,}
                    & {\rm for}\ \theta_{\rm h}=\theta \cr
                    {\displaystyle {1\over\sqrt{\theta_{\rm h}^2-\theta^2\,}}
                    \log\left({\theta_{\rm h}\over\theta}
                    +\sqrt{{\theta_{\rm h}^2\over\theta^2}-1\,}\right)\,,}
                    & {\rm for}\ \theta_{\rm h}>\theta
                   }\right.
        \end{eqnarray}
        and $\theta_{\rm h}=r_{\rm h}/D_{\rm L}$. $\theta_{\rm E}$ and $\theta_0$ are given by Equation~(\ref{thetaEtheta0}) with
        $M$ being the total mass of the lens.

        The time delay (difference) equation is given by (cf. Equation~(\ref{GRtimedelay}))
        \begin{eqnarray}\label{HernTimeDelayEq}
          && {\displaystyle{{\tilde D}_{\rm LS}\over{\tilde D}_{\rm L}{\tilde D}_{\rm S}}{H_0\Delta t\over(1+z_{\rm L})}=
          {(\theta_++\theta_-)\over(g_++g_-)}\left[q-{1\over 2}(\theta_++\theta_-)(g_+-g_-)\right]} \cr
          && \quad\quad\quad\quad\quad\quad {\displaystyle +\theta_{\rm E}^2\left[p-{q(f_++f_-)\over(g_++g_-)}
          +{(\theta_++\theta_-)(g_+f_--g_-f_+)\over(g_++g_-)}\right]}\,,
        \end{eqnarray}
        where
        \begin{eqnarray}\label{pq}
          && {\displaystyle p(\theta_+,\theta_-,\theta_{\rm h})=\log\left(\theta_+\over\theta_-\right)
          +\theta_{\rm h}\left[h(\theta_+,\theta_{\rm h})-h(\theta_-,\theta_{\rm h})\right]\,,} \\
          && {\displaystyle q(\theta_+,\theta_-,\theta_{\rm h})=
          \left(\theta_+^2-\theta_{\rm h}^2\right)h(\theta_+,\theta_{\rm h})
          -\left(\theta_-^2-\theta_{\rm h}^2\right)h(\theta_-,\theta_{\rm h})
          -\theta_{\rm h}\log\left(\theta_+\over\theta_-\right) \,.}
        \end{eqnarray}

        Eliminating $M$ from Equations~(\ref{HernLensEq}) \& (\ref{HernTimeDelayEq}), we obtain a relation between $\mathfrak{a}_0$ and $H_0$
        in terms of observed quantities $\theta_\pm$, $\theta_{\rm h}$, $\Delta t$.
        Explicitly,
        \begin{equation}\label{a0H0}
          {4\tilde{D}_{\rm L}\tilde{D}_{\rm LS}\over\tilde{D}_{\rm S}}{a_0\over cH_0}
          ={\left[(\theta_++\theta_-){\cal D}-(f_++f_-){\cal N}\right]^2
          \over(g_++g_-)^2{\cal D}{\cal N}}\,,
        \end{equation}
        and $M$ is given by
        \begin{equation}\label{Hernmass}
          {4\tilde{D}_{\rm LS}\over\tilde{D}_{\rm L}\tilde{D}_{\rm S}}{GH_0M\over c^3}
          ={{\cal N}\over{\cal D}}\,,
        \end{equation}
        where
        \begin{equation}\label{num}
          {\cal N}={\tilde{D}_{\rm LS}H_0\delta T\over\tilde{D}_{\rm L}\tilde{D}_{\rm S}(1+z_{\rm L})}
          -\,{(\theta_++\theta_-)\over(g_++g_-)}\left[q-{1\over 2}(\theta_++\theta_-)(g_+-g_-)\right]\,,
        \end{equation}
        \begin{equation}\label{den}
          {\cal D}=p-{q(f_++f_-)\over(g_++g_-)}+{(\theta_++\theta_-)(g_+f_--g_-f_+)\over(g_++g_-)}\,.
        \end{equation}
        For a given $H_0$, $\mathfrak{a}_0$ is given explicitly by Equation~(\ref{a0H0}).
        For a given $\mathfrak{a}_0$, Equation~(\ref{a0H0}) is a third order equation in $H_0$.
        It can be shown that only one solution is suitable.

        For other interpolation functions, we still have the lens equation and the time delay equation,
        but the mass $M$, $\mathfrak{a}_0$, $H_0$ cannot be separated as nicely as for the Bekenstein form.
        In any case, there are two relations for three variables.
        If the mass can be estimated from other methods, say velocity dispersion of the lensing galaxy,
        then it is possible to get $\mathfrak{a}_0$ and $H_0$ simultaneously.


\end{document}